\documentclass[final,5p,times,twocolumn]{elsarticle}

\usepackage{amssymb}
\usepackage{amsmath}

\usepackage{graphicx}
\usepackage{caption}
\usepackage{float}
\usepackage{algorithm}
\usepackage{algpseudocode}
\usepackage{amsfonts}
\usepackage{hyperref}
\usepackage{listings}
\usepackage{xcolor}
\usepackage{url}
\usepackage{natbib}
\usepackage{lineno}

\lstdefinestyle{jsonstyle}{
    backgroundcolor=\color{white},   
    basicstyle=\small\ttfamily,
    breaklines=true,                 
    keywordstyle=\color{blue},       
    commentstyle=\color{green},      
    stringstyle=\color{red}          
}

\begin{document}

\begin{frontmatter}



\title{Semantic Model for the SKA Regional Centre Network} 



\author[1]{Edgar Ribeiro João\corref{cor1}} 
\ead{edgarj@iaa.es} 
\author[1]{Manuel Parra-Royón} 
\ead{mparra@iaa.es} 
\author[1]{Julián Garrido\corref{cor1}} 
\ead{jgarrido@iaa.csic.es}

\affiliation[1]{organization={Department of Extragalatic Astronomy, Instituto de Astrofísica de Andalucía (CSIC)},
            addressline={Glorieta de la Astronomía, s/n}, 
            city={Granada},
            postcode={18008}, 
            country={Spain}}

\cortext[cor1]{Corresponding author}

\begin{abstract}
The unprecedented volume of data from the Square Kilometre Array (SKA) telescopes will require the implementation of robust and solid strategies for efficient data processing and management. In this context, the SKA Regional Centre Network (SRCNet)— a collaborative global infrastructure comprising multiple regional centres distributed across various geographical regions around the globe—is poised to play a critical role. This network will be instrumental in facilitating the effective handling and analysis of extensive data streams generated by the telescopes, thereby enabling significant advancements in astronomical research and exploration. This paper introduces a semantic model implemented with JSON-LD designed specifically for the SRCNet, detailing its architecture, data distribution, and computing service. By explicitly defining nodes, resources, relationships, and workflows, this model lays a foundation for interoperability and efficient resource management within the distributed network. The model presented in this text supports two possible configurations: centralized and decentralized— depending where data reside—, enabling a future service broker to efficiently plan workflows by querying nodes for real-time system availability. Consistency tests conducted using SPARQL queries were made on the model in order to validate and test its integrity. Therefore, this research contributes to the advancement of semantic modeling in astronomy by addressing the semantic model for the SRCNet, a topic that has not been previously explored. This semantic model serves as a precursor to the development of a precise mathematical representation of the network and establishes a foundational framework for a future service broker.
\end{abstract}




\end{frontmatter}




\section{Introduction}
\label{sec:introduction}

The Square Kilometre Array (SKA) telescopes will deliver substantial advances in sensitivity, resolution, and survey speed \cite{schilizzi2024square}. The array comprises two primary components, namely SKA-Mid in South Africa and SKA-Low in Western Australia, covering a wide frequency range from 50~MHz to 14~GHz \cite{kramer2015pulsar, garrido2023status}. The scientific objectives of the SKA are broad, spanning studies of the Cosmic Dawn and the origins of life, among many others \cite{an2019science, aaska14}. Given the immense volume of data ($~$700 PB per year) to be transferred, processed, stored, and disseminated globally, the SKA is widely regarded as one of the most demanding challenges in astrophysics and big data. The community is already acquiring significant experience in processing and analysing such data through SKA precursors and participation in the SKA Data Challenges \cite{bonaldi2021square, hartley2023ska, bonaldi2025square}.

To cope with this unprecedented data scale, the SKA Observatory (SKAO) will rely on a worldwide network of SKA Regional Centres (SRCNet) \cite{SKADeliveryPlan}. The SRCNet is being designed to efficiently archive, distribute, and process the scientific data products generated by the telescopes, while tailoring data processing capabilities to the specific requirements of the users \cite{bolton2019ska, srcnetDocs}. In practice, the scientific community will access SKAO data through the SRCNet, which will also provide resources for data analysis, processing  and visualisation. The development of this federated network is catalysing research and innovation in areas such as science gateways and distributed computing \cite{bertocco2023, parra2024bringing}.

A key step in enabling the SRCNet is the ability to explicitly describe its architecture: the network of interconnected nodes, their computing, storage, and networking resources, and the services available to end scientist users. Such descriptions must be machine-actionable to support efficient operations, interoperability between SRCNet nodes, and the integration of advanced functionalities such as AI-driven brokering and workflow orchestration, among others.

Several specification approaches are available within the Cloud Computing (CC) domain that can inform this effort. Cloud Modelling Languages (CMLs) such as TOSCA \cite{binz2013opentosca}, CloudML \cite{bergmayr2015evolution}, OCCI \cite{edmonds2011open}, and CAMEL \cite{rossini2017cloud} provide declarative mechanisms to describe infrastructure resources, nodes, services, policies, and relationships. These languages are commonly used to define application topologies as collections of software components and their resource requirements, with the goal of ensuring that the delivered quality meets user expectations \cite{korontanis2024survey}. Complementary to these modelling approaches, structured data vocabularies such as Schema.org \cite{patel2014analyzing} or Microdata \cite{nogales2016linking} are widely adopted for defining types, properties, and attributes in a programmatic and understandable manner, with applications ranging from web annotation to database interoperability \cite{8454999}.  

Among the available serialisation formats, JavaScript Object Notation for Linked Data (JSON-LD) \cite{jsonld11, rfc2026} has become a preferred option for structuring data \cite{navarrete2019analyzing, navarrete2017use}. JSON-LD provides a JSON-based mechanism for representing Linked Data, enabling straightforward integration with existing web technologies and reusing vocabularies such as Schema.org\cite{mozSchema2024}, among others. 

Building on these foundations, this work proposes a semantic model for the SRCNet. The model is designed to represent the architecture, data distribution, and computing services of the network, enabling explicit descriptions of nodes, resources, relationships, and workflows. In doing so, it lays the groundwork for interoperability, efficient resource management, and advanced services across the distributed SRCNet infrastructure. 

The remainder of this paper is organised as follows. Section~\ref{sec:related works} provides an overview of related work on semantic and ontological modeling. Section~\ref{sec:semantic model} presents the JSON-LD–based semantic model for representing the SRCNet and its implementation. Section~\ref{sec:validation} describes the deployment of the proposed model in a semantic database and its validation through SPARQL queries. Finally, Section~\ref{sec:final considerations and future works} offers concluding remarks and outlines future work.

\section{Related Work}
\label{sec:related works}

The literature underscores the importance of knowledge structure and management across various domains. In the field of Data Analytics, Bandera et al. \cite{bandara2020semantic} address challenges in engineering Data Analytics Solutions (DAS) arising from the heterogeneity of software components. This research emphasizes the necessity of ontologies as a formal basis for semantic modeling and analyses how semantic models represent diverse areas of knowledge related to DAS.

Parra-Rayon et al. \cite{parra2020semantics} propose a unified framework aimed at standardizing the definitions and descriptions of Data Mining services. This framework tackles management aspects, encompassing pricing strategies, authentication processes, and the nuances of Service Level Agreements, while also encapsulating the intricate technical components that constitute Data Mining workflows as services. The study introduces a semantic scheme for these definitions, laying the groundwork for broader standardization and industrialization of Data Mining services in cloud environments. Additionally, La-Ongsri et al. \cite{la2015incorporating} present a mechanism for incorporating ontologies in conceptual modeling techniques, demonstrating how this integration enhances the modeling of the semantic domain in database applications.

The previous works highlight the role of ontology as a foundational element for designing semantic models. Cao et al. \cite{cao2022concept} accentuate the critical function of semantic descriptions in indicating potential linguistic connections between objects, which are crucial for transferring knowledge across relationships and identifying novel connections. The authors propose a Concept-Enhanced Relation Network designed to facilitate video visual relation inference by integrating retrieved concepts with the local semantics of objects through a gating mechanism, thereby generating concept-enhanced semantic representations. Continuing this discourse on semantic frameworks, Choudhary et al. \cite{choudhary2002text} introduce a method for generating feature vectors utilizing semantic relations between words in a sentence, capturing semantic relations through the Universal Networking Language (UNL), a recently proposed semantic representation for sentences.

The issue of semantic interoperability using ontologies in distributed and federated data processing is also referenced by Vesely et al. \cite{vesely2004cern}, where the authors argue that the current trend in grey literature management is gravitating toward institutional repositories built on distributed and federated models.

Within the context of semantic technologies, there has been significant growth in research related to Big Earth Data. Narock et al. \cite{narock2017semantics} underscore the formal limitations of the Semantic Web. They provide an oceanographic example to illustrate these limitations, highlighting challenges surrounding the semantic encoding of observations and the use of semantics during analysis. Potential solutions to each challenge are presented, showing that comprehensive application of semantic technologies is immensely beneficial for Big Earth Data. It is important to note that the Semantic Web is characterized, according to Wu et al. \cite{wu2014semantic}, as an extension of the current web designed to facilitate not only human consumption of information but also computational processing of large-scale data. This study introduces key Semantic Web technologies, provides detailed analysis of how they address the heterogeneous variety of life sciences big data, and highlights the role of Semantic Web technologies in offering solutions in the era of big data. These assertions are substantiated by Narock et al. \cite{narock2013crowdsourcing}, where the authors confirm that semantic technologies effectively address data integration challenges related to data size, variety, and complexity. Their research examines Big Data within geosciences, posing essential questions regarding the integration of crowdsourcing and semantics. Thus, we conclude that multiple challenges related to volume, velocity, variety, and variability exist across different domains associated with big data. These challenges encompass storage, search, transfer, sharing, analysis, processing, visualisation, and semantics \cite{khan2013addressing}.

In the context of astronomy, ontologies, vocabularies, and metadata schemas are particularly relevant as they enable the semantic description of data and mapping of underlying concepts. The Data Management Infrastructure in astronomy \cite{CECCONI2025100991} is organized around the International Virtual Observatory Alliance (IVOA) \cite{quinn2004international}, which prioritizes interoperability. The IVOA defines schemas, protocols, and vocabularies, which are produced and maintained by the community through various working groups (WGs), such as the Data Models WG, Data Access Layer WG, and Semantics WG \cite{ivoaSemantics2024, CECCONI2025100991}. The IVOA regularly issues recommendations and standards to promote interoperability, including vocabularies employed in other reference documents \cite{gray2023units,derriere2011ivoa}. Technically, the IVOA defines vocabulary types for informal knowledge organization and strict hierarchies of classes and properties, while also allowing the use of IVOA vocabularies without specific RDF tooling \cite{demleitner2023vocabularies}.

Taking into account the discussions in this section and the introduction, the semantic model for the SKA Regional Centre Network (SRCNet) aims to enhance interoperability and resource orchestration. This model addresses dimensions that the International Virtual Observatory Alliance (IVOA) has not yet covered as of the time of writing. By implementing such a framework, SRCNet will contribute to the unified management and utilization of astronomical data, reinforcing collaboration and facilitating more efficient data sharing among international stakeholders in the astronomy community. This initiative promises not only to streamline processes but also to foster innovation through improved access to information and analytical capabilities, ultimately advancing research and discovery in the field of astronomy.

\section{Semantic Model}
\label{sec:semantic model}

Semantic Model play a critical role in structuring and interpreting data in meaningful way. It act as a foundational framework encapsulating the relations and attributes associated with several entities within a system. Therefore, semantic model is completely significant in contexts where data interoperability and clarity are primordial, such as in databases, knowledge, and information retrieval systems \cite{cao2022concept}.

In this work, the comprehensive representation of the SRCNet is defined through a semantic model aiming to describe all its components, dependencies, entities, attributes and the relations established among them. The architecture of the SRCNet \cite{salgadoska} shows how nodes of computing, storage, networking and services interact and participate in the global computing system.

\captionsetup{justification=centering}
\begin{figure*}[htbp]
    \centering
    \includegraphics[width=0.9\textwidth]{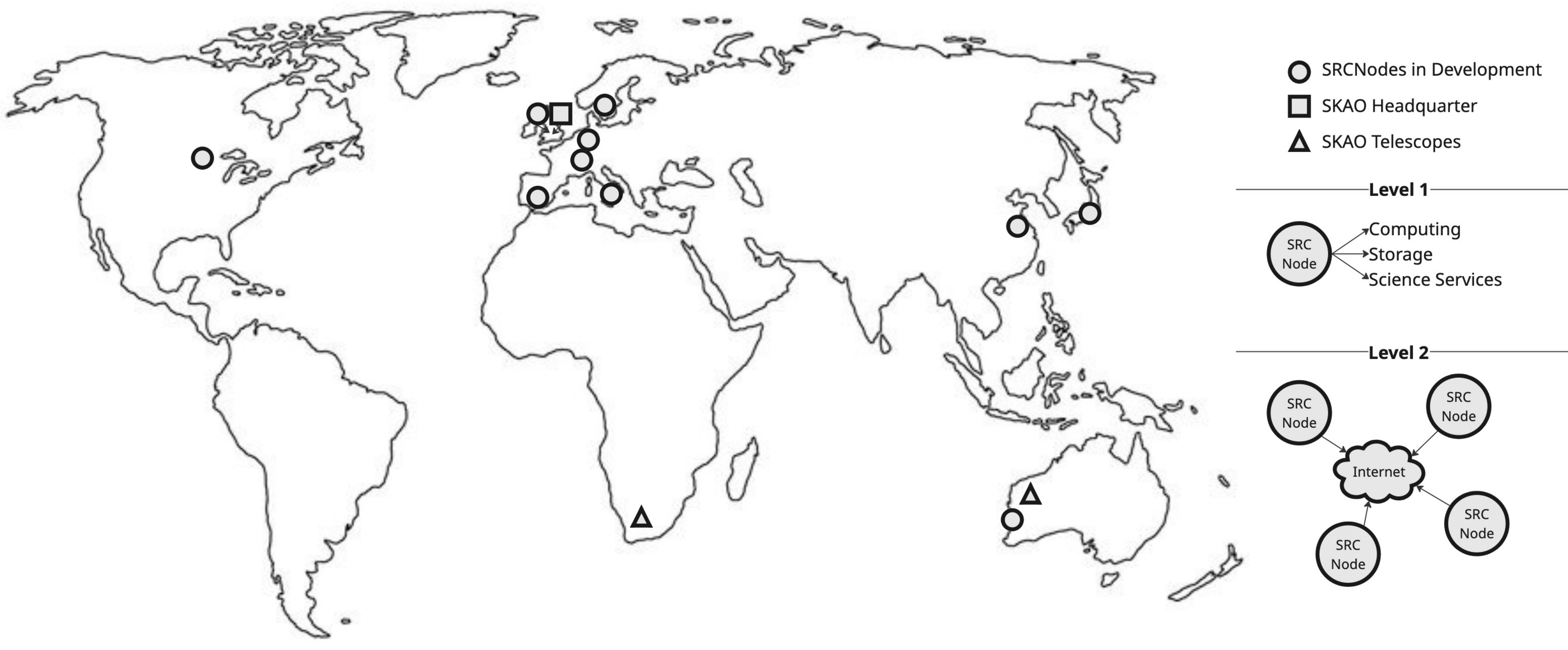}
    \caption{Geographical distribution of the current SRCNet nodes at the time of writing, as well as the location of SKAO telescopes and SKAO headquarters. On the right, each component provided by each SRCNet node is also described at a general level (computing, storage, and science services). The interconnection between the SRC nodes is carried out via the Internet.}
    \label{fig:srcnet-info}
\end{figure*}

The current description of the SRCNet model is structured around two levels. Level 1 refers to the interconnection of SRCs through the internet and level 2 corresponds to each SRC as a local node within a country, equipped with a large computing infrastructure that is internally interconnected, while also linking to the rest of the SRCNet nodes via the internet, as shown in the Figure \ref{fig:srcnet-info}.

\subsection{Description of the SRCNet features}
\label{subsec:description of the SRCNet features}

This section presents the key features of the SRCNet in terms of connectivity, science services, data topology and distribution, data access and locality, and data movement and workflow costs.  

Regarding to connectivity, each node within the SRCNet is interconnected with all other nodes via the Internet, shaping a  distributed network topology. Connectivity is a critical aspect of the infrastructure, as it directly affects data transfer efficiency, interoperability, and the orchestration of distributed scientific workflows. The following properties characterise the inter-node connections:

\begin{itemize}
    \item \textbf{Bandwidth} ($Mbps$ \/ $Gbps$): Defines the maximum data transmission capacity of a network link. Higher bandwidth enables larger data volumes to be transferred simultaneously, directly influencing throughput and the efficiency of large-scale data movement.
    
    \item \textbf{Latency} ($ms$): Represents the time elapsed for a data packet to travel from one SRC node to another. Low latency is essential for interactive or time-sensitive operations, such as data visualisation or remote workflow execution.
    
    \item \textbf{Jitter} ($ms$): Indicates the variation in latency between transmitted packets. Low jitter ensures predictable communication timing, which is particularly important for real-time applications such as remote instrument control or data streaming services.
    
    \item \textbf{Packet Loss} (\%): Expresses the proportion of data packets lost during transmission. Packet loss above acceptable thresholds can significantly degrade the quality of services and should be monitored to maintain data integrity and reliability.
    
    \item \textbf{Availability} (\% uptime): Reflects the proportion of time during which a network connection is operational and stable. High availability (typically $>$ 99.9\%) is a fundamental requirement for maintaining continuous data access and service reliability across the SRCNet.
\end{itemize}

Concerning to science services, each SRC provides a set of web-based services that enable user interaction, data access, and remote execution capabilities \footnote{See SRCNet Services Operator: \url{https://ska-telescope.gitlab.io/src/kb/ska-src-docs-operator/}}. The behaviour and quality of these services significantly influence user experience and overall system reliability. The following measurable characteristics should be explicitly captured in the semantic model:

    \begin{itemize}
        \item \textbf{Response Time} ($ms$): The time it takes for a web service to process a request and return a response. This is a crucial indicator of the speed of a service.
        \item \textbf{Load Time} ($s$): The time required for a web service content to load completely on the client. This includes server response time and data transfer time.
        \item \textbf{Scalability} (number of concurrent requests \/ elasticity ratio): The ability of a web service to handle an increase in the number of concurrent requests without degrading performance. This may include the ability to load balance and distribute requests to multiple servers.
        \item \textbf{Availability} (\% uptime): The proportion of the time that a web service is operational and accessible. High availability is essential to ensure that services are always accessible to users.
        \item \textbf{Performance Under Load} (transactions per second \/ response degradation ratio): How a web service performs when subjected to high levels of traffic or load. Includes stress and load tests to measure performance under extreme conditions
        \item \textbf{Server Resource Utilisation} (\% CPU \/ \% Memory \/ Network Throughput in $Mbps$): The amount of server resources, such as CPU, memory and bandwidth, used by a web service. Resource efficiency is crucial to maintaining optimal performance.
     
    \end{itemize}

In relation to data topology and distribution, scientific data products are asymmetrically distributed across multiple regional nodes. Each piece of data---for instance, a measurement set or calibrated image---may exist as one or more replicas across different SRCs. This asymmetric replication strategy reflects both physical constraints and optimisation policies of the network. The following elements are considered:

    \begin{itemize}
        
        \item \textbf{Replication Factor} (number of replicas): Each data entity may have one or multiple replicas. A low replication factor (e.g., 1) often indicates large file size or limited storage capacity, whereas a higher factor enhances redundancy and data availability.

        \item \textbf{Replica Placement} (SRC identifier): Defines the specific SRCs hosting each replica. Not all SRCs maintain the same datasets, resulting in heterogeneous data coverage across the network.
    
        \item \textbf{Local Data Dependency}: Services operating at each SRC depend on the presence or remote accessibility of required datasets. When data cannot be transferred due to bandwidth or replication constraints, computations must be executed at the node where the data resides.

        \item \textbf{Data Granularity} (collection size / number of datasets): A data product can represent a single file or a collection of related observations. Collections may be heterogeneously distributed---for example, five subsets in espSRC, three in sweSRC, and two in itSRC.
        
        \item \textbf{Data and Service Placement}:The services can perform specific operations on the data products such as visualisation or a data analysis pipeline (e.g. to extract sources from the data). The data can be locally available in the SRC where the user is executing the full workflow, but the data can also be located in different SRCs. Any workflow or pipeline should be analysed in advance to provide a set of metrics regarding data access and dependencies to execute the workflow successfully. Through this analysis, it is possible to determine what penalties may arise from data movements or the use of external computing resources and services outside the local SRC of the user.
    \end{itemize}

With respect to data access and locality, the performance of data access will depend on many factors, some of the most critical are:

    \begin{itemize}
        \item \textbf{External} ($I/O$) \textbf{performance} ($MB$\/$s$): Throughput achieved during data transfer operations between the SRC and external systems. Higher $I/O$ rates reduce data staging and retrieval times.
        \item \textbf{Replica Proximity} (network hops \/ geographical distance): Represents the relative distance between SRCs hosting the same data. Closer replicas generally enable faster access and lower transfer latency.
        \item \textbf{Connection Availability} (\% uptime): Indicates the operational state of the link between SRCs or storage nodes. High availability is critical for maintaining reliable access paths.
        \item \textbf{Internal Bandwidth} ($Gbps$): Measures the intra-SRC network capacity available for transferring data within local infrastructure components (e.g., between compute and storage nodes).
        \item \textbf{Network Latency} ($ms$): The end-to-end time delay in data transmission, as previously defined. Lower latency improves data streaming and workflow synchronisation.
        
    \end{itemize}

Regarding to data movement and workflow costs, they are related to the costs and resources required when executing certain operations involving data movement and workflows. Here are described some of these costs:

    \begin{itemize}
        \item \textbf{Data Transfer Cost} ($Watts$ \/ $GB transferred$ \/ latency penalty): The cost of moving a piece of data from one SRC to another impacts directly in execution time, energy consumption, bandwidth and latency.
        \item \textbf{Workflow Execution Cost} (CPU hours / GPU hours / energy consumption): The cost of executing a workflow that involves both local and external data is associated to energy consumption, bandwidth and storage capacity.
        \item \textbf{Storage Cost} (\ $euro$\/$TB$\/$month$): Eventual penalties applied to the data include cost of storing data which is directly related to the amount of data to be stored.
        \item \textbf{Data Latency}: Eventual movement of the data to the local SRC which affects data download times.
        \item \textbf{Data Transfer Costs}: Costs related to bandwidth, latency, data egress and data ingress. 
        \item \textbf{Time of overall processing}: Includes costs for virtual machines, serverless functions, or dedicated hardware.
        \item \textbf{Data processing on the external SRC}: Includes costs associated with computing resources such as CPU, GPU, memory, and others. 
    \end{itemize}

A decomposition into different nodes of the full graph, hosted in the repository \cite{edgar202517608322} associated to this work, is provided as a visual representation of the semantic model for the SRCNet. This illustrates the relationships between entities and attributes, highlights the importance of organizing and structuring data to improve understanding and support effective data management, and at the same time delivers accurate and detailed information about its orchestration.

The set of characteristics described above provides the foundation for constructing a semantic representation of the SRCNet. These features capture both the structural and operational aspects of the network—ranging from connectivity and computing to data topology, access, and cost metrics. Based on these properties, our semantic model must distinguish between two complementary components: a \textit{static model}, which defines the persistent entities, resources, and relationships that describe the SRCNet architecture; and a \textit{dynamic model}, which represents time-dependent properties such as performance metrics, data flows, and resource utilisation. Next subsections will cover both models.

\subsection{Static model of the SRCNet}
\label{subsec:static model of the SRCNet}
\subsubsection{SRC Identification}
\label{subsubsec:SRC Identification}

\begin{figure*}[htbp]
    \centering
    \includegraphics[width=0.9\textwidth]{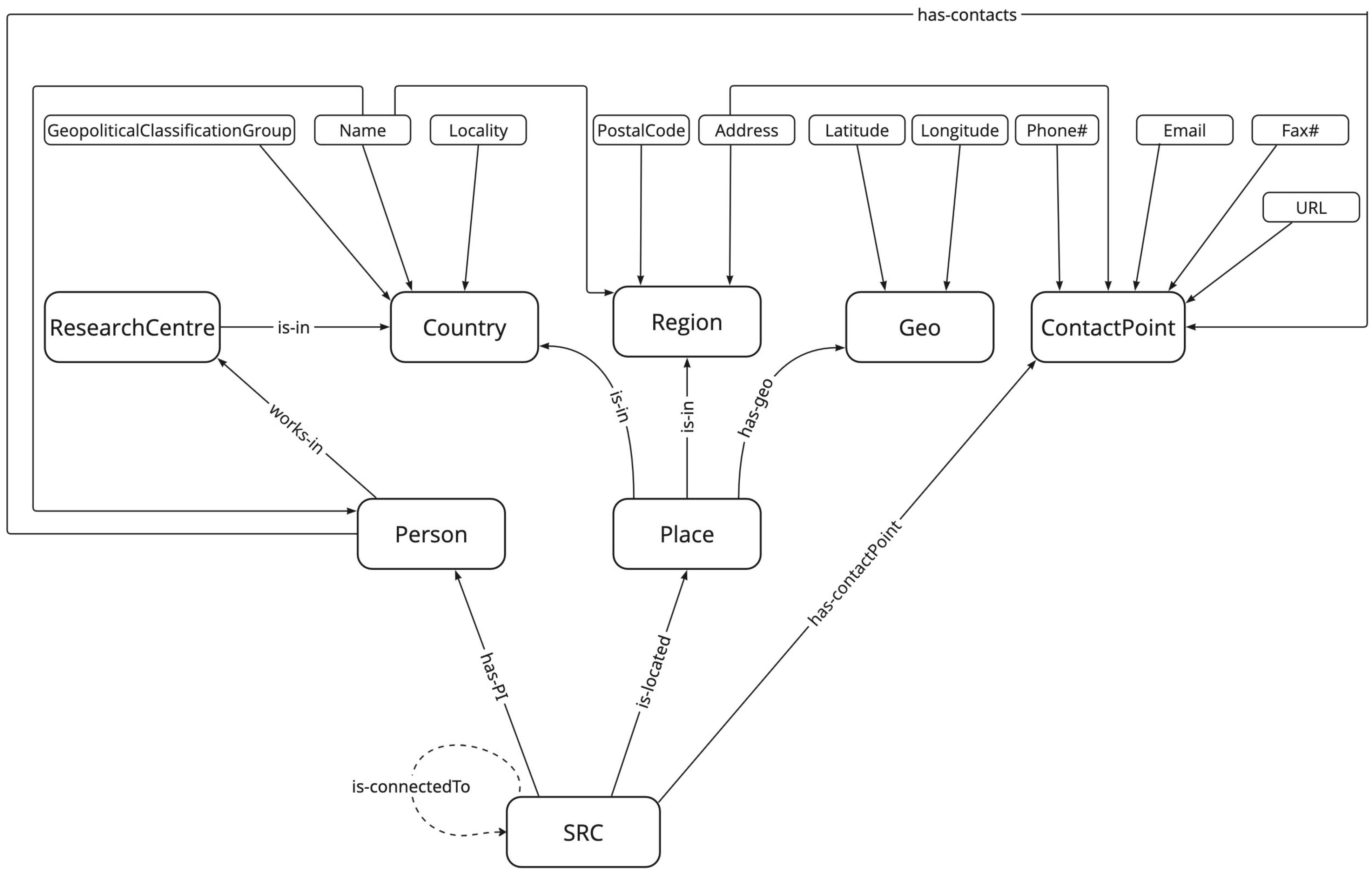}
    \caption{Semantic model of SRCNet entities showing spatial hierarchy (Country–Region–Place), institutional affiliations (ResearchCentre–Person–PI), and interconnections among SRCs, including geographic and contact information relationships.}
    \label{fig:srcnode-identification}
\end{figure*}

 Figure \ref{fig:srcnode-identification} presents a detailed description of the SRC, aiming to provide all the information related to its identification and location, associated entities, properties, attributes, and relations.

\captionsetup{justification=centering}

The SRCNet consists of several SRCs. At the time of writing, each SKAO member country or prospective country is leading an initiative to create an SRC. This entity would allow to represent each of those national SRC initiatives. In general, an SRC is associated to a region or country and this element is connected to other specific entities that would provide more detailed information. In particular, the principal investigator (PI) of the SRC and the place where the SRC is located are defined through the relations \texttt{has-PI} and \texttt{is-located}, respectively.

\begin{itemize}

    \item \textit{Person}: Defined through the relationship \texttt{has-PI}, \textit{Person} represents individual entity, encapsulating personal information of the maximum responsible of the SRC. The affiliation of \textit{Person} is described through the relation \texttt{works-in} as \textit{ResearchCentre}, representing an academic or research institution, in a particular country, where \textit{Person} works.

    \item \textit{Place}: Represents a specific location that may be relevant to the SRC (e.g. headquarters or major provider), linking through defined relations to the following entities:

        \begin{itemize}
            \item \textit{Country}: The main country where the SRC is. It includes attributes like name, locality and the geopolitical classification group.
            \item \textit{Region}: The region where the SRC belongs. It includes attributes like name of the region, postal code and address. In the event of multinational SRCs (SRC supported by several countries), Region would allow to describe that the SRC has a geographical scope larger than a country. 
            \item \textit{Geo}: The geo coordinates where the SRC is located. It includes attributes like latitude and longitude.
        \end{itemize}

    \item \textit{ContactPoint}: A person or team working in the SRC management. It includes attributes such as phone number, email, fax number, address and url.
   
 \end{itemize}

The proposed model defines the entity \textit{Node} as a resource provider linked to the SRC through the relation \texttt{is-composedBy}, specifying that an SRC might consist of one or several \textit{Node} entities, as will be discussed ahead \ref{fig:services-node}.  At the time of writing, both cases coexist within the SRCNet. A node can be described in terms of storage, compute, network, security policies, and services, playing each of them a specific role in the system.

\subsubsection{Storage}

\begin{figure}[htbp]
    \centering
    \includegraphics[width=0.5\textwidth]{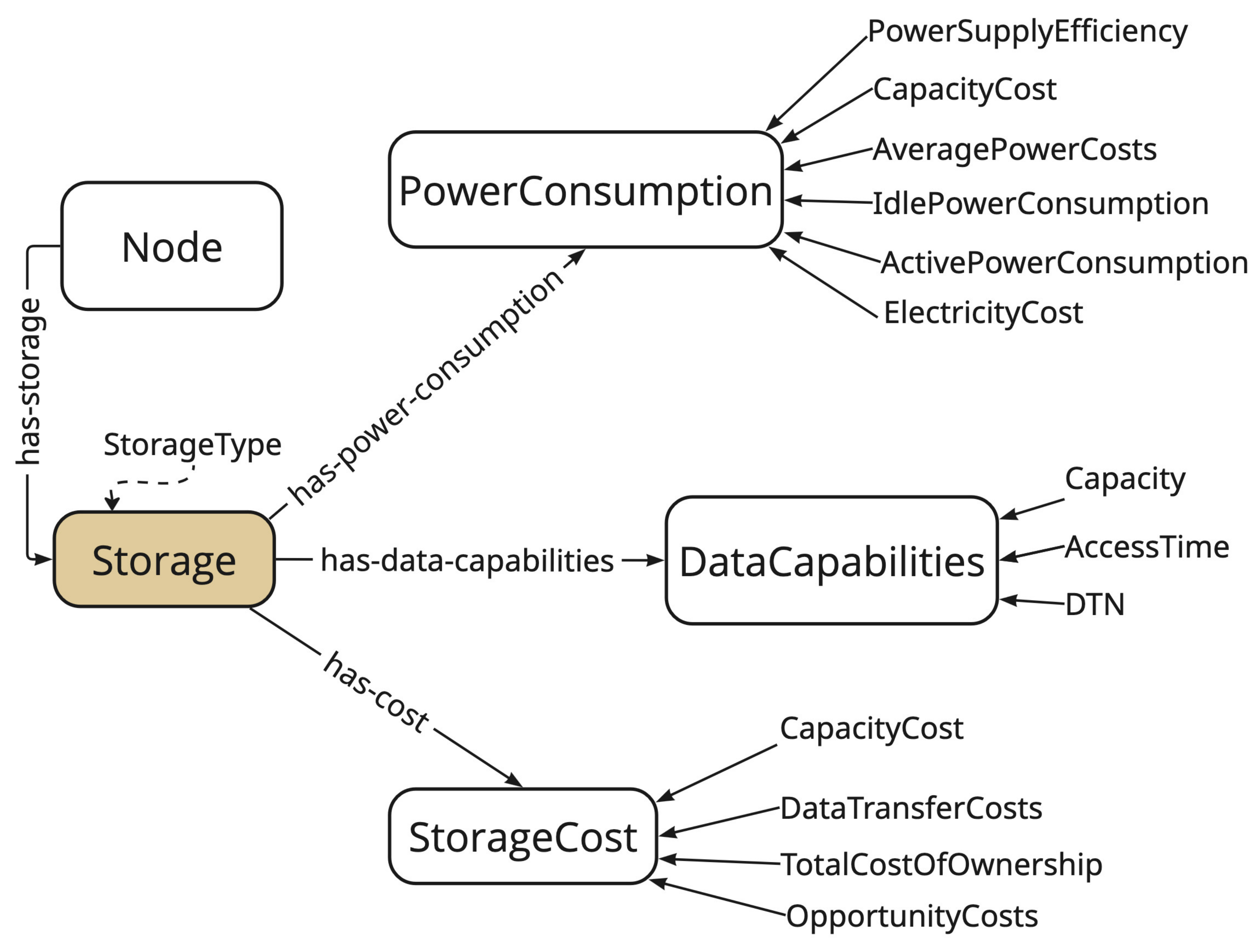}
    \caption{Detailed view of entities related to the storage node}
    \label{fig:storage-node}
\end{figure}

This is a component of the \textit{Node} responsible for storing, manage, and retrieve data within a distributed or centralized storage architecture. It plays a critical role in data storage systems, specifically in cloud computing, distributed files systems, and network-attached storage environments. It includes entities like \textit{PowerConsumption}, \textit{DataCapabilities} and \textit{Cost}, as shown in the Figure \ref{fig:storage-node}.

\begin{itemize}
    \item \textit{PowerConsumption}: Represents the theoretical electrical energy consumed by the hardware components, storage devices, processors, memory, and network interfaces during operation. This entity includes \textit{PowerSupplyEfficiency}, \textit{CapacityCost}, \textit{AveragePowerCosts}, \textit{IdlePowerConsumption}, \textit{ActivePowerConsumption}, and \textit{ElectricityCost}. 
    \item \textit{DataCapabilities}: Refers to the range of functionalities allowing the node to handle data operations, including data storage, retrieval, processing, and management. It includes attributes like  \textit{Capacity}, \textit{AccessTime}, and the presence of \textit{DTN} (Data Transfer Node). 
    \item \textit{StorageCost}: Incorporates all expenses incurred throughout the lifecycle of a storage entity, including ongoing operational costs which are critical for comprehensive financial analysis. This entity contains attributes such as \textit{CapacityCost}, \textit{DataTransferCosts} \textit{TotalCostOfOwnership}, and \textit{OpportunityCosts}. The total cost of ownership encompasses all financial expenditures associated with the storage, providing a robust framework for informed investment decisions. Capacity cost quantifies the economic implications of scaling storage capacity, which is essential for effective budgeting in the context of increasing data volumes. Data transfer costs delineate the expenses associated with data movement, highlighting the necessity for efficient workflows to mitigate unnecessary financial burdens. Finally, the opportunity costs assess the potential benefits forfeited when selecting one storage solution over another, thereby facilitating strategic decision-making. 
\end{itemize}

\captionsetup{justification=centering}

\subsubsection{Computing}

\begin{figure}[htbp]
    \centering
    \includegraphics[width=0.5\textwidth]{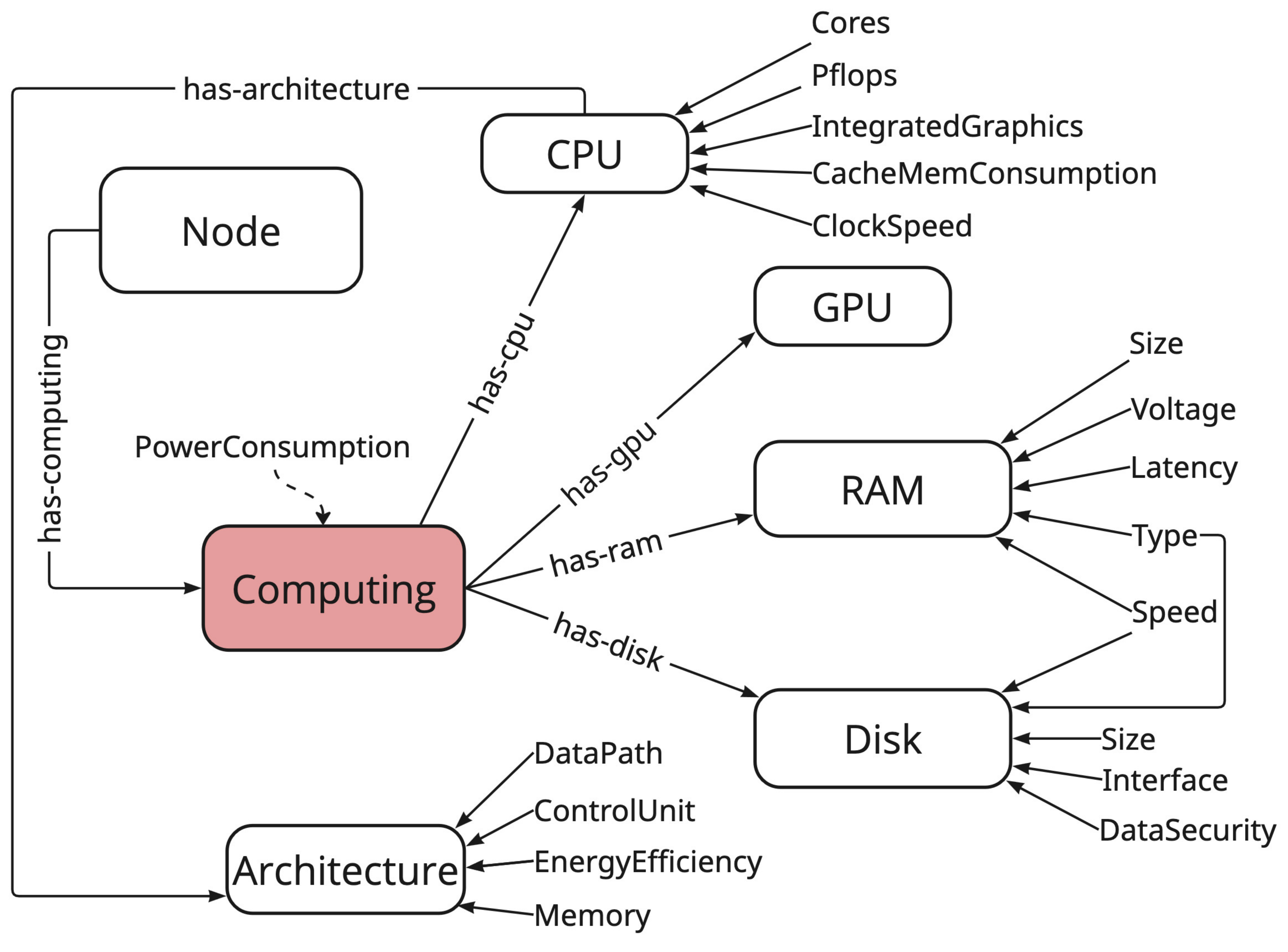}
    \caption{Overview of the computing entity highlighting its key components, namely \textit{CPU}, \textit{GPU},\textit{RAM}, \textit{Disk} and their corresponding attributes.}
    \label{fig:compute-node}
\end{figure}

This entity describes processing components such as \textit{CPU}, \textit{GPU}, \textit{RAM}, \textit{Disk}, and \textit{Architecture} performing computations on data, operating as a part of a larger distributed computing environment where it collaborates with all other nodes of the \textit{SRC} to handle specific tasks. Figure \ref{fig:compute-node}illustrates the connections to the computing entity components and indicates their respective attributes.  

\begin{itemize}
    \item \textit{CPU} is the primary component of the compute node responsible for executing instructions, performing operations and manage data flow allowing the overall functionality of the computing system.  It describes the computational capacity and features, in terms of CPU. Linked to this component is defined, through the relationship \texttt{has-architecture}, the entity \textit{Architecture} which is the conceptual design and fundamental operational structure of a computer system, related to the organization of the hardware components, their interactions and the principles leading their operation. It includes attributes such as \textit{DataPath}, \textit{ControlUnit}, \textit{EnergyEfficiency}, and \textit{Memory}. 
    
    \item \textit{GPU}
    
    \item \textit{RAM} is the component used to storage temporally data and instructions that CPU needs to execute tasks efficiently. This entity refers to the total RAM capacity in the \textit{Node}.  
    
    \item \textit{Disk} describes the physical medium utilized to store and retrieve data. While \textit{Storage} refers to a broader term that encompasses various methods and technologies for storing and managing data within a network system.
    
    \textit{Disk} includes attributes such as \textit{type} (HDD or SSD), \textit{size}, \textit{interface}, \textit{DataSecurity} and \textit{speed}.  
\end{itemize}

\captionsetup{justification=centering}

\subsubsection{Network} 
\begin{figure}[htbp]
    \centering
    \includegraphics[width=0.5\textwidth]{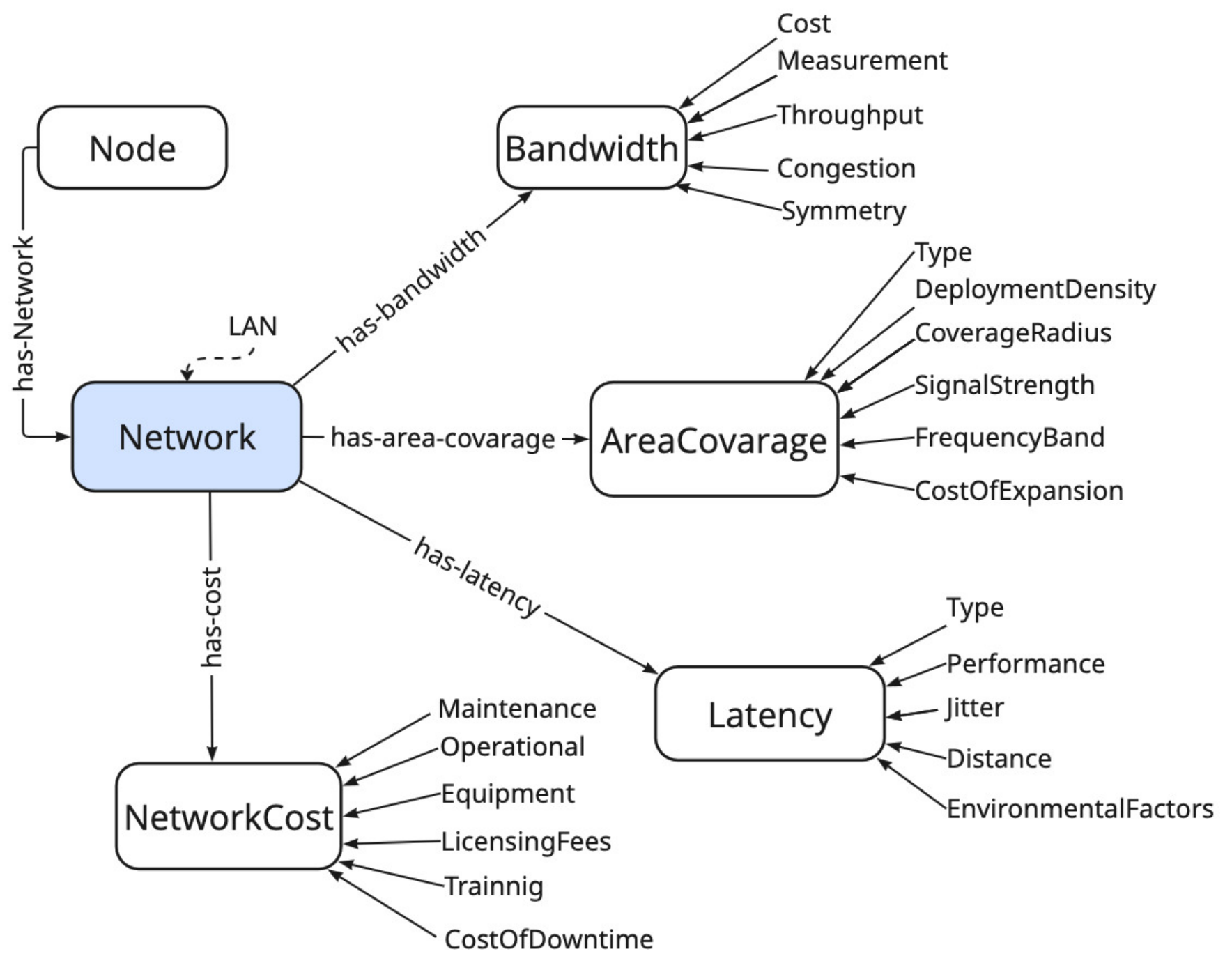}
    \caption{The illustration of network entity components describing their relationships and attributes}
    \label{fig:network-node}
\end{figure}

Defined as key unit that facilitates the communication and data exchange in the system, the \textit{Network} is the entity with the ability to send, receive, process data and interact with all other nodes in a structured way. Figure \ref{fig:network-node} shows the Network components: \textit{Bandwidth}, \textit{AreaCoverage}, \textit{Latency}, \textit{NetworkCost}, and their corresponding attributes.

\captionsetup{justification=centering}

\begin{itemize}
    \item \textit{Bandwidth}: Refers to the maximum rate at which data are transmitted over a communication channel within the network. This entity  can be considered as a parameter that  determines the capacity of the node to face data traffic, being measured in bits per second, kilobits per second, megabits per second or gigabits per second. It includes attributes like \textit{Measurement}, \textit{Throughput}, \textit{Congestion}, and \textit{symmetry}, and \textit{Cost}.
    \item \textit{AreaCoverage}: Represents the geographical range in which the node can transmit and receive data signals and determines the ability of the node to provide connectivity and service to users or devices within a specific area. It includes attributes such as \textit{DeploymentDensity}, \textit{CoverageRadius}, \textit{FrequencyBand}, \textit{CostOfExpansion}, among others.
    \item \textit{Latency}: Refers to the time of delay experienced in the transmission of data from one point to another within a network. This is a critical performance metric that affects the responsiveness and efficiency of the network communications. \textit{Latency} can include attributes such as \textit{Type}, \textit{Performance}, \textit{Jitter}, \textit{Distance}, \textit{EnvironmentalFactors}, among others.
    \item \textit{NetworkCost} is a critical component that involves several financial aspects related to the deployment, operation, and maintenance. Includes costs of \textit{Maintenance}, \textit{Operational}, \textit{LicensingFees}, \textit{Training}, \textit{CostOfDowntimes}, among others.
\end{itemize}

\subsubsection{Energy efficiency and sustainability}

\begin{figure}[htbp]
    \centering
    \includegraphics[width=0.5\textwidth]{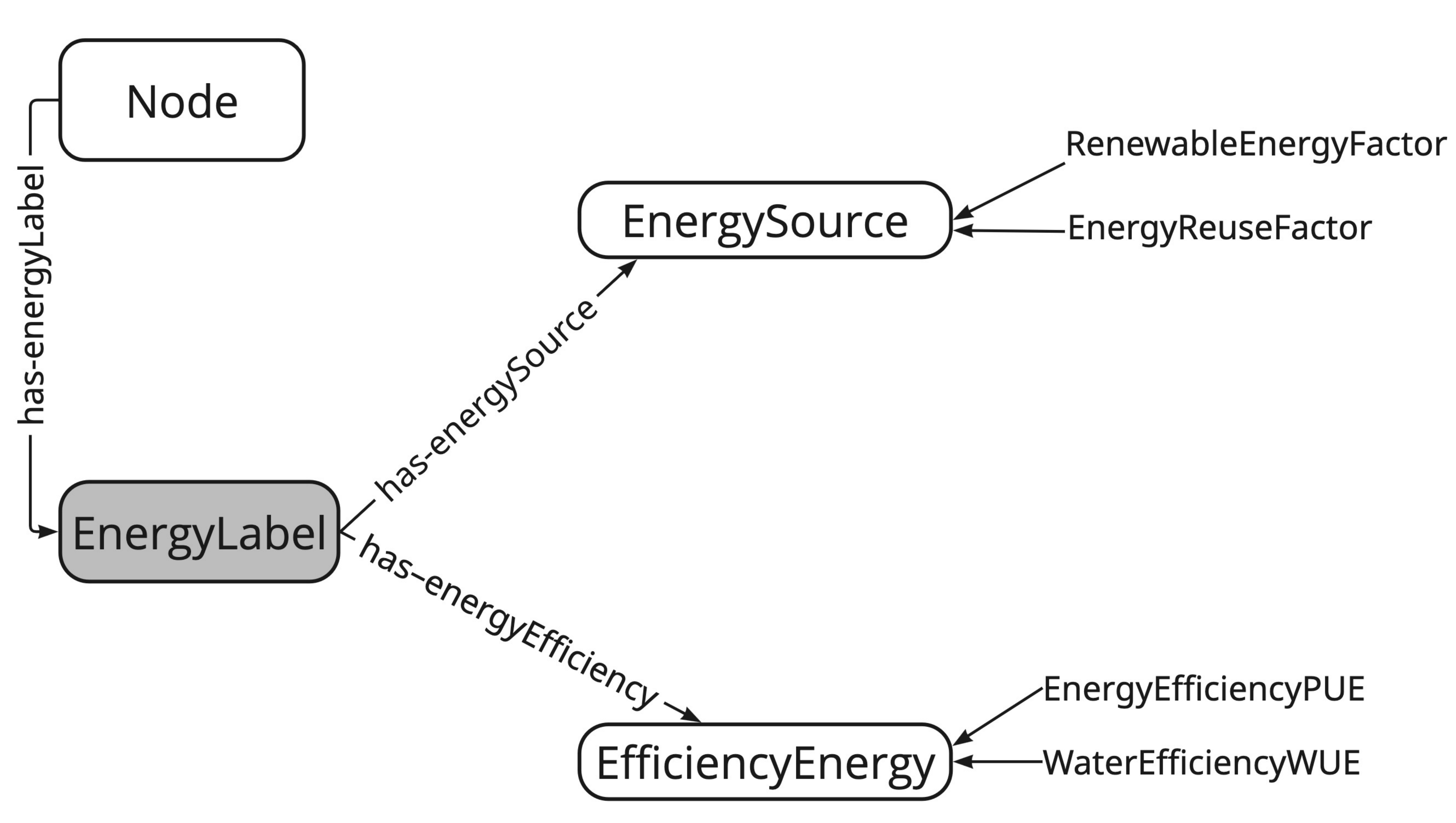}
    \caption{This illustration depicts the components of the energy labeling system, showcasing categories such as \textit{EnergySource} and \textit{EfficiencyEnergy}}
    \label{fig:energy-node}
\end{figure}

As part of the energy efficiency and sustainability of the data centres, different initiatives aim to design energy label \cite{kontinakis2023eu, eu2025datacentres}. For instance, the Energy Efficiency Directive driven by the European Commission for the Energy, Climate Change and Environment defends that an efficient use of energy will contribute to reduce the overall energy consumption in European countries ensuring the goal of reducing greenhouse gas emissions by at least 55 percent, compared to 1990, until 2030 \cite{eu2023eed}. Figure \ref{fig:energy-node} shows that the SRCNet model includes an entity defined as \textit{EnergyLabel} through the relationship \texttt{has-energyLabel} connected from \textit{Node}.

\captionsetup{justification=centering}

The entity \textit{EnergyLabel} captures important aspects associated to the origin of energy what is directly related to the upstream sustainability impacts of data centre operations, the efficient use of energy responding to the question about how efficient are the energy resources used in the data centre, and circularity measures identifying possible sustainability measures and the downstream impacts. This entity is composed by the following components:

\begin{itemize}

    \item \textit{EnergySource} is focused on the origin of the energy and it is generally considered as the most important metric climate protection during the operational phase including aspects related to the carbon free sources, besides power and the source of water. \textit{EnergySource}, defined through the relationship \texttt{has-energySource}, possesses metrics such as \textit{RenewableEnergyFactor} and the \textit{EnergyReuseFactor}, defined by the European Commission \cite{eu2025datacentres} as:
    
    \begin{itemize}

    \item  \textit{RenewableEnergyFactor} (REF) is the ratio of total renewable energy consumption($E_{RES - TOT}$) to total energy consumption(${E_{DC}}$).

    \begin{equation}
        REF = \frac{E_{RES - TOT}}{E_{DC}}
    \end{equation}
    
    \item \textit{EnergyReuseFactor} (ERF): ratio of waste heat reuse ($E_{REUSE}$) to total energy consumption ($E_{DC}$).

    \begin{equation}
        ERF = \frac{E_{REUSE}}{E_{DC}}
    \end{equation}

\end{itemize}
    
    \item \textit{EfficiencyEnergy} is a critical component of sustainability, emphasizing the need to optimize energy use while maintaining performance levels. By focusing on metrics such as \textit{EnergyEfficiencyPUE} that represents a key performance indicator used to assess the energy efficiency of data centres and \textit{WaterEfficiencyWUE} indicating the metric used to evaluate the efficiency of water use in data centres, particularly regarding cooling systems quantifying the amount of water used per unit of energy consumed by the IT equipment and serving as an important indicator of the sustainability and environmental impact of data centre operations. It is defined in this model through the relationship \texttt{has-energyEfficiency} containing two metrics formally defined as following \cite{eu2025datacentres}:
    
    \begin{itemize}
    
    \item Power Usage Effectiveness (PUE): ratio of total energy consumption ($E_{DC}$) to total energy consumption of IT equipment ($E_{IT}$).
    
    \begin{equation}
        PUE = \frac{E_{DC}}{E_{IT}}
    \end{equation}

    \item Water Usage Effectiveness (WUE): ratio of total water input ($W_{IN}$) to total energy consumption of IT equipment ($E_{IT}$).

    \begin{equation}
        WUE = \frac{W_{IN}}{E_{IT}}
    \end{equation}

    \end{itemize}

\end{itemize}

\subsubsection{SRCNet Science Services} 
\begin{figure*}[htbp]
    \centering
    \includegraphics[width=1\textwidth]{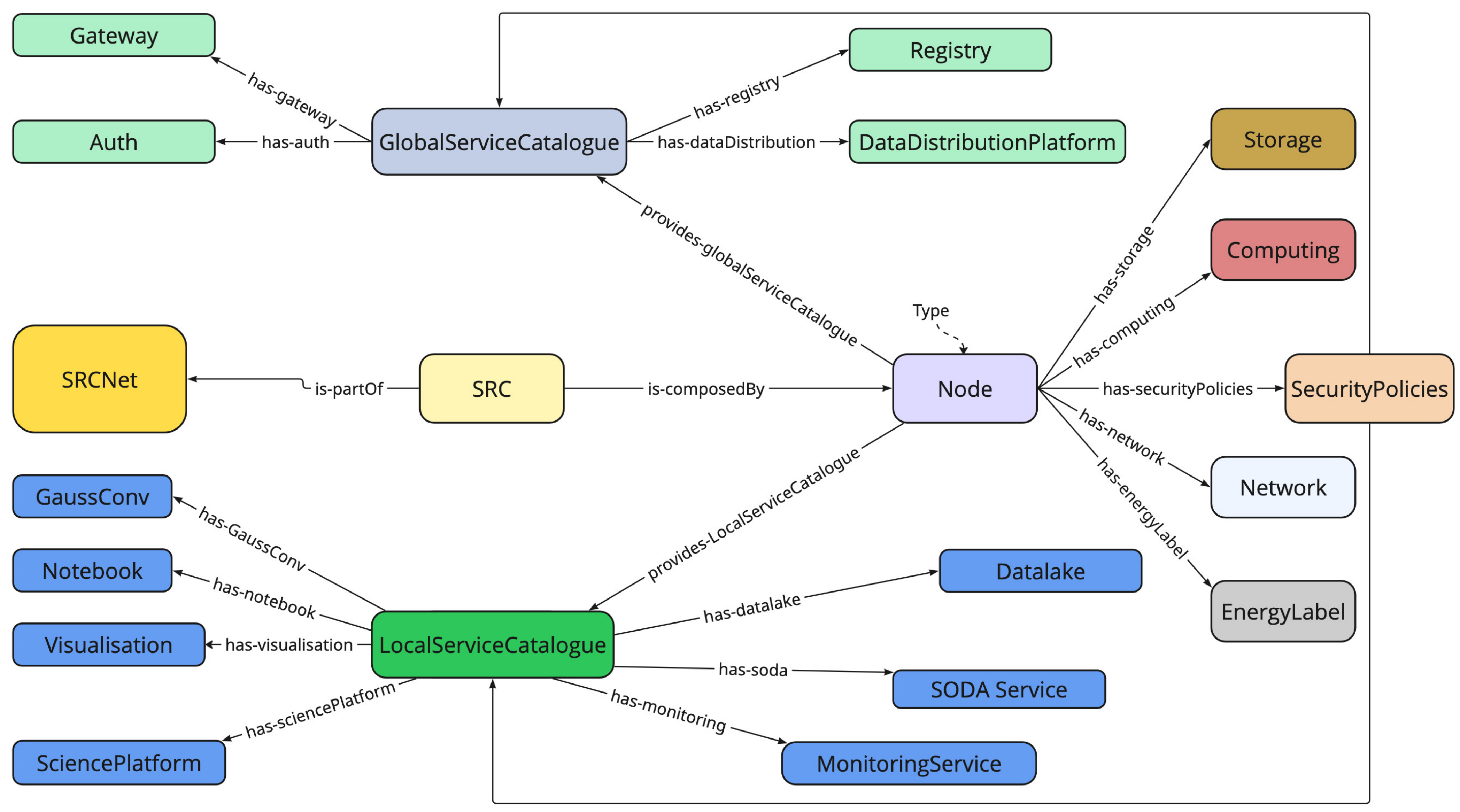}
    \caption{A simplified representation of the SRCNet model illustrating how SRC as part of the SRCNet is composed by different nodes}
    \label{fig:services-node}
\end{figure*}

In the context of the SRCNet, services are specialized entities providing specific functionalities to the other nodes in the system. They act as intermediary that processes requests, performs tasks, and delivers outputs, facilitating interaction and communication among several nodes of the system. There are specific services oriented to the scientific community providing them functionalities such as scientific analyses, data visualisation, data access, data manipulation, among others. Two types of services are defined in the SRCNet: the first type, Global Services refer to centralised services that are needed for the overall system.  To this type of services is defined the entity \textit{GlobalServiceCatalogue}, connected to \textit{Node}, through the relationship \texttt{has-globalServiceCatalogue} cantaining services such as, \textit{ScienceGateway}, \textit{Auth}, \textit{Registry}, and \textit{Data DistributionPlatform} \cite{SKAOReport0009}. The second type, Local Services represents the services \cite{Barisits2019} locally provided by the SRC such as \textit{Notebook}, \textit{Visualisation}, \textit{SciencePlatform}, \textit{MonitoringService}, \textit{Datalake}, and \textit{SODAservice}. To this type of services is defined the entity \textit{LocalServiceCatalogue}, also connected to \textit{Node}, through the relationship \texttt{has-localServiceCatalogue} where each service has the following components: \texttt{infrastructure}, \texttt{PodConfiguration}, \texttt{storage}, \texttt{NetworkingAndAccessibility}, and dependencies with their specific attributes, as we can visualize in the full diagram \cite{edgar202517608322} of the model. Complementarily, a simplified view of the whole model is represented in Figure \ref{fig:services-node}, where the main relationships between the top level entities are depicted i.e. connecting all the entities described in previous figures \ref{fig:storage-node}, \ref{fig:compute-node}, \ref{fig:network-node}.

\captionsetup{justification=centering}

\textbf{Global Services:}

\begin{itemize}
    \item \textit{ScienceGateway}: A user interface layer in the SRCNet architecture providing access to several functionalities such that data discovery, visualisation, workflow creation, and execution. It is supported by the Gateway Pattern, where declared services dynamically configure user interface elements, API, methods, and server components ensuring the scalability, extensibility, and a consistent user experience across SRCNet nodes \cite{bertocco2023}.
    \item \textit{Auth}: The authentication in the SRCNet is implemented like a federated authorisation infrastructure. It includes national federations through an international inter-federation service like eduGain, enabling users to authenticate using institutional credentials. This system connects these credentials to a centrally coordinated unique SRCNet identity, enabling a secure access to the data, computing, and others provided resources. 
    \item \textit{Registry}: Refers to the metadata management system allowing query, discovery, and management of metadata related to science entities, user details, and Datasets. It guarantees secure access, translation, and execution of queries to support functionalities such as data discovery, table uploads, and metadata updates.  
    \item \textit{DataDistributionPlatform}: In the SRCNet, Data Distribution has to do with the ingestion and replication of science data produced by SKA telescopes into the SRCNet data lake and derived data products. The system certifies efficient data logistics through the distribution of primary and secondary copies of data across SRCNet nodes based on factors such as resources, network, connectivity, and user location. It uses hierarchical storage management (HSM) enabling the optimisation of storage layers for performance and cost. 
\end{itemize}

\textbf{Local Services}:

\begin{itemize}
    \item \textit{Notebook}: It is a scientific analysis tool that provides the appropriate environment to create and share documents containing live codes and equations. We can mention web-based notebooks like Jupyter \cite{kluyver2016jupyter, granger2021jupyter}, Google Colab, Kaggle Kernels, Microsoft Azure, IBM Watson Studio, Deepnote, RStudio Cloud, and Zeppelin \cite{cheng2018building} used to analize logs stored in HDFS. These notebooks essentially allow users to explore and analyse data in a very interactive way.  
    \item \textit{Visualisation}: It can be defined as a tool or set of tools used to provide graphical representations of Data ensuring a better understanding and analysis. An example of these tools is Cube Analysis and Rendering Tools for Astronomy (CARTA) \cite{comrie2021carta}, a service that provides tools to visualize data cubes, enabling astronomers to explore multi-dimensional datasets interactively \cite{labadie20253d}. 
    \item \textit{SciencePlatform}: This service is used to support astronomers by providing a range of services, tools for data access, storage, analysis, desktops, jobs, visualisation, and notebook. 
    \item \textit{Monitoring Service}: This service monitors applications performance using tools such as  Prometheus plus Grafana for metrics, Loki plus Fluentd for log management, and AlertManager for notifications.
    \item \textit{Datalake}: This service acts as an abstraction layer allowing Rucio or other equivalent deployment to interact with several types of storage systems such as local disk storage, tape storage, cloud storage, or other data repositories. For instance, the use of HDFS for long-term storage and backup of logs and data could be considered a form of datalake where large volumes of structured and unstructured data are stored for analysis.
    \item \textit{SODA}: A framework focused on the integration and management of data services in a service-oriented architecture, emphasizing the use of data as a service enabling users to access and manipulate data through standardized interfaces and protocols. SODA is an example of service that would facilitate data analysis. In particular, SODA services produce new cut-out data cubes. Complementarily, to data analysis services would contribute to develop a catalogue that would facilitate the creation workflows and analysis pipelines. It is particularly crucial in environments like the SRCNet where data need to be shared across different nodes.
    \item \textit{GaussConv}: Is a function in the same line of the SODA, as Function as a Service (FaaS), that can be deployed and run for the science users.
\end{itemize}

\subsection{Dynamic model of the SRCNet}
\label{subsec:dynamic model of the SRCNet}

Considering that in the SRCNet resources allocation can change over the time, a model that explains how the dynamical aspects involved communicate and interact each other focusing on the system run-time behaviour, is presented. Essentially, in this subsection, we distinguish the dynamic from the statical variables looking to those parts of the graph that may change over the time because of the dynamic nature of the system.

The Figure \ref{fig:Dynamic} shows how the \textit{SRC}, as part of the SRCNet through the relation \texttt{is-composedBy} is linked to \textit{Node} and how \textit{Node}, in turn, through the relation \texttt{has-nodeAvailability} is connected to \textit{NodeAvailability} containing the following entities:

\begin{itemize}
   
\item \textit{ComputingAvailability}: Refers to the degree to which computing resources are available and accessible in any time \textit{t}. This entity possesses parameters such as \textit{AvailabilityCPUcores}, \textit{UsedCPUcores}, \textit{AvailabilityGPUcores} and \textit{UsedGPUcores}.

The \textit{AvailabilityCPUcores} represent the number of CPU cores available that can be utilized at any time \textit{t}, while \textit{UsedCPUcores} indicates how many of those cores are currently engaged in processing tasks. This relationship is vital, as a higher number of used CPU cores can lead to increased latency and reduced responsiveness if the available cores are insufficient to handle the workload.
Similarly, the \textit{AvailableGPUcores} and \textit{UsedGPUcores} reflect the capacity and utilization of graphical processing units, which are essential for tasks requiring parallel processing, such as machine learning and graphics rendering. The interaction between these two sets of cores is significant; for instance, if the \textit{UsedGPUcores} are nearing their limit while the \textit{AvailableGPUcores} are low, it can lead to bottlenecks in performance, affecting the ability of the system to process data efficiently. Moreover, the balance between \textit{CPU} and \textit{GPU} core usage is critical in optimizing resource allocation. If the \textit{CPU} is heavily utilized while the \textit{GPU} remains underused, it may indicate a need for workload redistribution to enhance overall performance. Conversely, if both \textit{CPU} and \textit{GPU} resources are fully utilized, it may signal the need for scaling up resources or optimizing the workload to maintain computing availability. 

\item \textit{StorageAvailability}: Essential for ensuring reliable access to data, \textit{StorageAvailability} refers to the degree to which data storage resources are accessible and operational in time \textit{t}. It contains metrics like \textit{AvailableCapacity} and \textit{UsedCapacity}.

\textit{AvailableCapacity} indicates the total storage space that can be utilized at any given moment, while \textit{UsedCapacity} represents the amount of storage that is currently occupied by data and applications. The relationship between these two metrics is crucial for effective data management and system performance.
When \textit{UsedCapacity} approaches the limits of \textit{AvailableCapacity}, the system may experience performance degradation, as insufficient storage can hinder the ability to save new data or execute applications that require additional space. This situation can lead to increased latency and potential data loss if the system is unable to accommodate new information.

\item \textit{NetworkAvailability}: Responsible for the connectivity level of the system in the time \textit{t}, this entity contains dynamic variables such as \textit{AvailableBandwidth} and \textit{UsedBandwidth}, criticals for evaluating the network performance and data transfer capabilities of a system. 

\textit{AvailableBandwidth} represents the maximum bandwidth that can be utilized for communication between nodes and external sources, while \textit{UsedBandwidth} indicates the current percentage of that bandwidth that is actively being consumed. The interaction between these two metrics is essential for ensuring efficient data flow and system responsiveness.
When \textit{UsedBandwidth} approaches the limits of \textit{AvailableBandwidth}, the system may experience congestion, leading to increased latency and reduced throughput. This congestion can significantly impact the performance of applications that rely on real-time data transfer, such as streaming services, online gaming, and cloud computing. If the bandwidth is fully utilized, it may result in packet loss or delays, adversely affecting user experience and application performance.
In addition, the relationship between bandwidth and other system resources, such as CPU and storage capacity, is significant. High \textit{UsedBandwidth} can strain processing resources, as data must be handled more frequently and quickly to keep up with the demands of active applications. Conversely, if the bandwidth is underutilized, it may indicate inefficiencies in data transfer protocols or application design, suggesting a need for optimization.

\item \textit{Dataset}: Expresses the availability of a file or set of files in the time \textit{t}. 

\item \textit{ServiceAvailability}: Encapsulates the operational status and accessibility of services running on a computing node. This entity is defined by several parameters that collectively provide insights into the performance, reliability, and user engagement of the services. \textit{ServiceName}, \textit{ServiceVersion}, \textit{ServiceType}, \textit{NumberOfUsers}, \textit{ActiveSession}, and \textit{Status} play a pivotal role in understanding the operational landscape of the system.

\textit{ServiceName} indicates the specific service currently running on the node, while \textit{ServiceVersion} denotes the version of that service. This information is crucial for maintaining compatibility and ensuring that users are accessing the most up-to-date features and security patches. The \textit{ServiceType} further categorizes the service, distinguishing between architectures such as microservices and monolithic applications. This classification can influence deployment strategies, scalability, and maintenance practices.
The \textit{NumberOfUsers} metric reflects the current user engagement with the service, providing insights into its popularity and demand. A high number of users can indicate a successful service but may also lead to resource strain if not managed properly. This is where the \textit{ActiveSession} metric becomes significant, as it indicates whether a process is currently engaged with the service and can take states such as active, idle, expired, or terminated. Monitoring active sessions helps in resource allocation and can inform decisions regarding scaling or optimizing service performance. The \textit{Status} refers whether service is running, stopped, or in an error state providing a snapshot of its operational health. A service in an error state may require immediate attention to prevent downtime and ensure user satisfaction. The interplay between these parameters is essential for maintaining a responsive and efficient computing environment, as they collectively inform decisions related to resource management, service optimization, and user experience.

\captionsetup{justification=centering}
\begin{figure*}[htbp]
    \centering
    \includegraphics[width=1\textwidth]{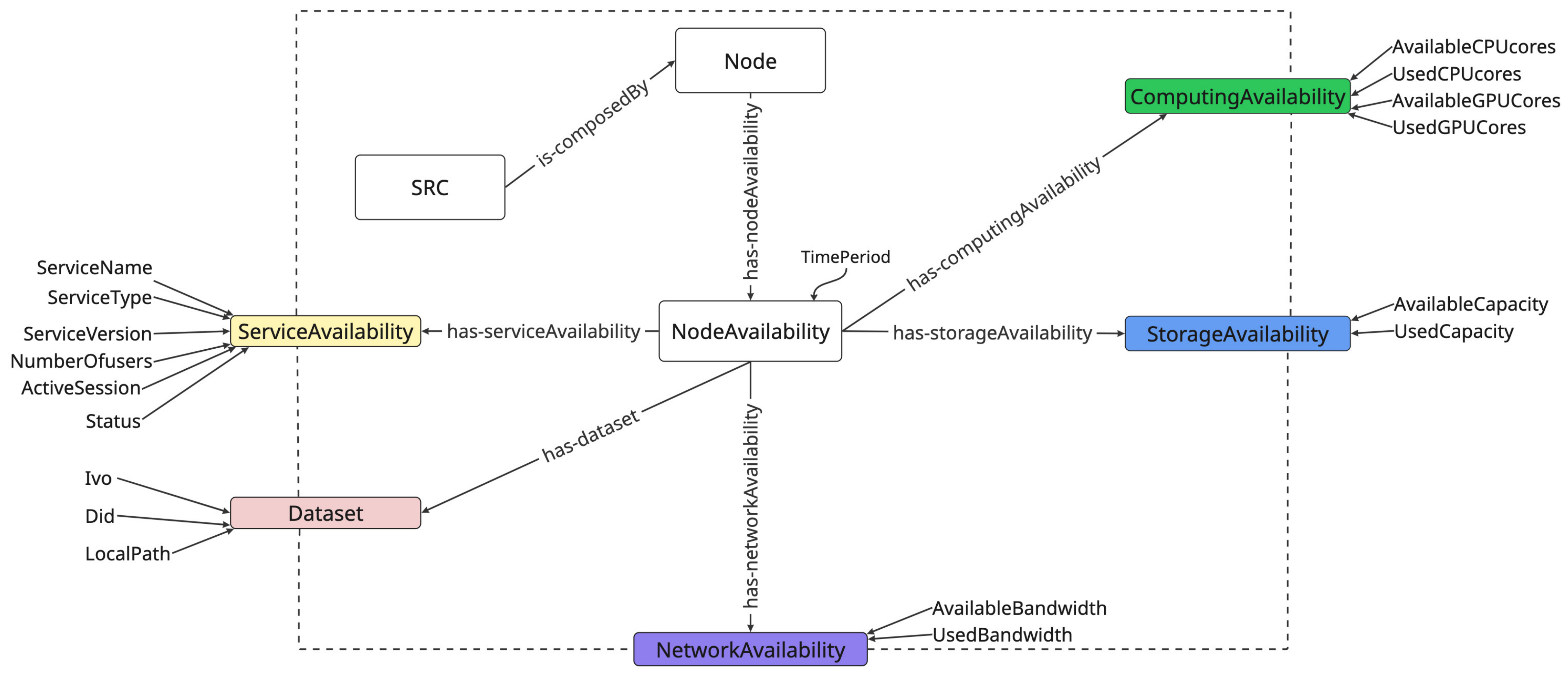}
    \caption{Diagram of the Dynamic Model: This illustration outlines the components and interactions within the SRCNet dynamic model, highlighting parameters and relationships.}
    \label{fig:Dynamic}
\end{figure*}

\item \textit{Connections between entities}: The diagram of the dynamic model exposed in the Figure \ref{fig:Dynamic}, illustrates the connections among entities of the model describing their corresponding dynamic variables.

Note that all entities within the system are interconnected, creating a complex web of interactions. \textit{ComputingAvailability}, which encompasses resources such as CPU cores and GPU cores, interacts closely with storage resources to access and process data. This interaction is facilitated through \textit{NetworkAvailability}, which serves as the communication backbone. The resources within \textit{ComputingAvailability} are responsible for activating the execution of services running on the node.
Conversely, StorageAvailability provides essential data to \textit{ComputingAvailability}, enabling effective data processing. This data transfer relies on the available network resources, which facilitate communication among the various entities. \textit{NetworkAvailability} not only enables connectivity for \textit{ServiceAvailability} but also supports the transfer of data, allowing services running on the node to interact with other entities or services seamlessly.
While \textit{ServiceAvailability} utilizes computing resources to execute workloads, it also accesses and stores data by leveraging the resources provided by \textit{StorageAvailability}. Furthermore, it communicates with other entities through the network resources, ensuring a cohesive and efficient operational environment. This intricate interplay among the entities underscores the importance of their interconnections in maintaining overall system performance and reliability.

\end{itemize}

\subsection{Model implementation}
\label{subsec:model implementation}

Based on the full description of the SRCNet structure, it is identified the cloud modeling language that best aligns with the specific features and requirements of the system.
It is important to refer that until this moment, we have presented the graph model through diagrams enabling a visual illustration of the semantic model. As previously exposed, this kind of applications are built using cloud modeling languages (CML), specifically, to describe and define the requirements of the application components in terms of resources. Therefore, according to the features and the architecture of the SRCNet \cite{salgadoska,parra2024bringing}, we have identified JSON-LD as the CML which better fits with the requirements of our system. Since that JSON-LD \cite{pyld2024} is designed as a portable syntax that can be used to express Linked Data. Moreover, JSON-LD is also useful when building interoperable Web Services and when storing Linked Data in JSON-based document storage engines. In the following subsections, we focus on presenting the most relevant components of the model rather than the comprehensive framework.

\subsubsection{SRCNet model implemented with JSON-LD}

The SRCNet model implemented through JSON-LD was performed in Python and the corresponding full JSON-LD file can be found in the repository \cite{edgar202517608322} created for this text. In the listings \ref{lst:cpu-definition}, \ref{lst:cpu-instantiation}, we present an example of how we have defined and instantiated a SRC node component.

\begin{lstlisting}[style=jsonstyle, caption={Definition of CPU component using JSON-LD},label={lst:cpu-definition}]

{
  "@context": {
    "@vocab": "http://schema.org/",
    "Computing": "http://example.org/CPU/Computing",
    "CPU": "http://example.org/CPU/",
    "cores": "http://example.org/CPU/cores",
    "pflops": "http://example.org/CPU/pflops",
    "clock": "http://example.org/CPU/clock",
    "qudt": "http://qudt.org/schema/qudt"
  },
  "@type": "Computing",
  "hasCPU": {
    "@type": "CPU",
    "cores": "Integer",
    "pflops":{
      "@type":"QuantityValue",
      "unit": "String",
      "value": "Float"
    },
    "clock": {
      "@type": "QuantityValue",
      "unit": "String",
      "value": "Float"
    }
  }
}
 \end{lstlisting}

In the example above, we define the component of a node type called \textit{CPU} with the attributes \textit{cores}, \textit{pflops}, and \textit{clock} and their corresponding given values. The context relates the defined terms to their corresponding URLs in the \url{schema.org} vocabulary.

Below, we create an instance \ref{lst:cpu-instantiation} for the \textit{CPU} node component defined above, including specific details of \textit{CPU} such as number of cores, number of pflops, and clock.

\begin{lstlisting}[style=jsonstyle, caption={Instantiation of CPU component},label={lst:cpu-instantiation}]

{
  "@context": {
    "@vocab": "http://schema.org/",
    "Computing": "http://example.org/CPU/Computing",
    "CPU": "http://example.org/CPU/",
    "cores": "http://example.org/CPU/cores",
    "pflops": "http://example.org/CPU/pflops",
    "clock": "http://example.org/CPU/clock",
    "qudt": "http://qudt.org/schema/qudt"
  },
  "@type": "Computing",
  "hasCPU": {
    "@type": "CPU",
    "cores": 240,
    "pflops":{
      "@type":"QuantityValue",
      "unit": "Petaflops",
      "value": 0.034
    },
    
    "clock": {
      "@type": "QuantityValue",
      "unit": "GHz",
      "value": 3.5
    }
  }  
}

 \end{lstlisting}

\subsubsection{Converting JSON-LD to RDF}

The conversion of JSON-LD into RDF was performed using \url{https://www.easyrdf.org/converter} aiming to inject the resulting RDF file graph in Apache-Jena Fuseki as a dataset. Considering the previous example, its corresponding RDF file is presented in the listing \ref{lst:rdf-format}.

\begin{lstlisting}[style=jsonstyle, caption={RDF format of CPU component},label={lst:rdf-format}]

<?xml version="1.0" encoding="UTF-8"?>
<rdf:RDF xmlns:rdf="http://www.w3.org/1999/02/22-rdf-syntax-ns#" xmlns:ns0="http://example.org/CPU/" xmlns:schema="http://schema.org/" xmlns:xsd="http://www.w3.org/2001/XMLSchema#">
  <ns0:Computing>
    <schema:hasCPU>
      <ns0: rdf:nodeID="g191352">
        <ns0:cores rdf:datatype="http://www.w3.org/2001/XMLSchema#integer">240</ns0:cores>
        <ns0:pflops>
          <schema:QuantityValue rdf:nodeID="g191360">
            <schema:unit>Petaflops</schema:unit>
            <schema:value rdf:datatype="http://www.w3.org/2001/XMLSchema#double">3.4E-2</schema:value>
          </schema:QuantityValue>
        </ns0:pflops>
        <ns0:clock>
          <schema:QuantityValue rdf:nodeID="g191368">
            <schema:unit>GHz</schema:unit>
            <schema:value rdf:datatype="http://www.w3.org/2001/XMLSchema#double">3.5E0</schema:value>
          </schema:QuantityValue>
        </ns0:clock>
      </ns0:>
    </schema:hasCPU>
  </ns0:Computing>
</rdf:RDF>

\end{lstlisting}

\section{Validation}
\label{sec:validation}

To deploy the proposed model we have instantiated it with some data and have uploaded it as RDF format in Apache Jena-Fuseki enabling to make SPARQL queries to test its consistency. In the following, we expose some of the SPARQL queries that can be made in our model. 

We can start by listing all \textit{SRCs} of the SRCNet with their respective countries, locations and localities, as shown in the query presented in listing \ref{lst:sparQL1}. However, its possible to include, in this query, other parameters such as the principal investigators indicating their contacts of e-mail, telephone, affiliations, SRC region if its applicable, the geographical coordinates and others.

\begin{lstlisting}[style=jsonstyle, caption={SPARQL query for listing the SRCs},label={lst:sparQL1}]

PREFIX SRCNet: <http://example.org/SRCNet>
PREFIX SRCNode: <http://example.org/SRCNode>
PREFIX schema: <http://schema.org/>


SELECT ?SRCNode_Name?country_name?SRCNode_location?Locality
WHERE {
  ?node a SRCNode: .
  ?node schema:name ?SRCNode_Name .
  
  ?node schema:location?location.
  ?location schema:isInAcountry?country.
  ?country schema:name?country_name.
  
  ?location schema:name?SRCNode_location.
  
  ?location schema:isInAcountry?country.
  ?country schema:Locality?Locality.
}
\end{lstlisting}

\begin{figure}[H]
    \centering
    \includegraphics[width=0.5\textwidth]{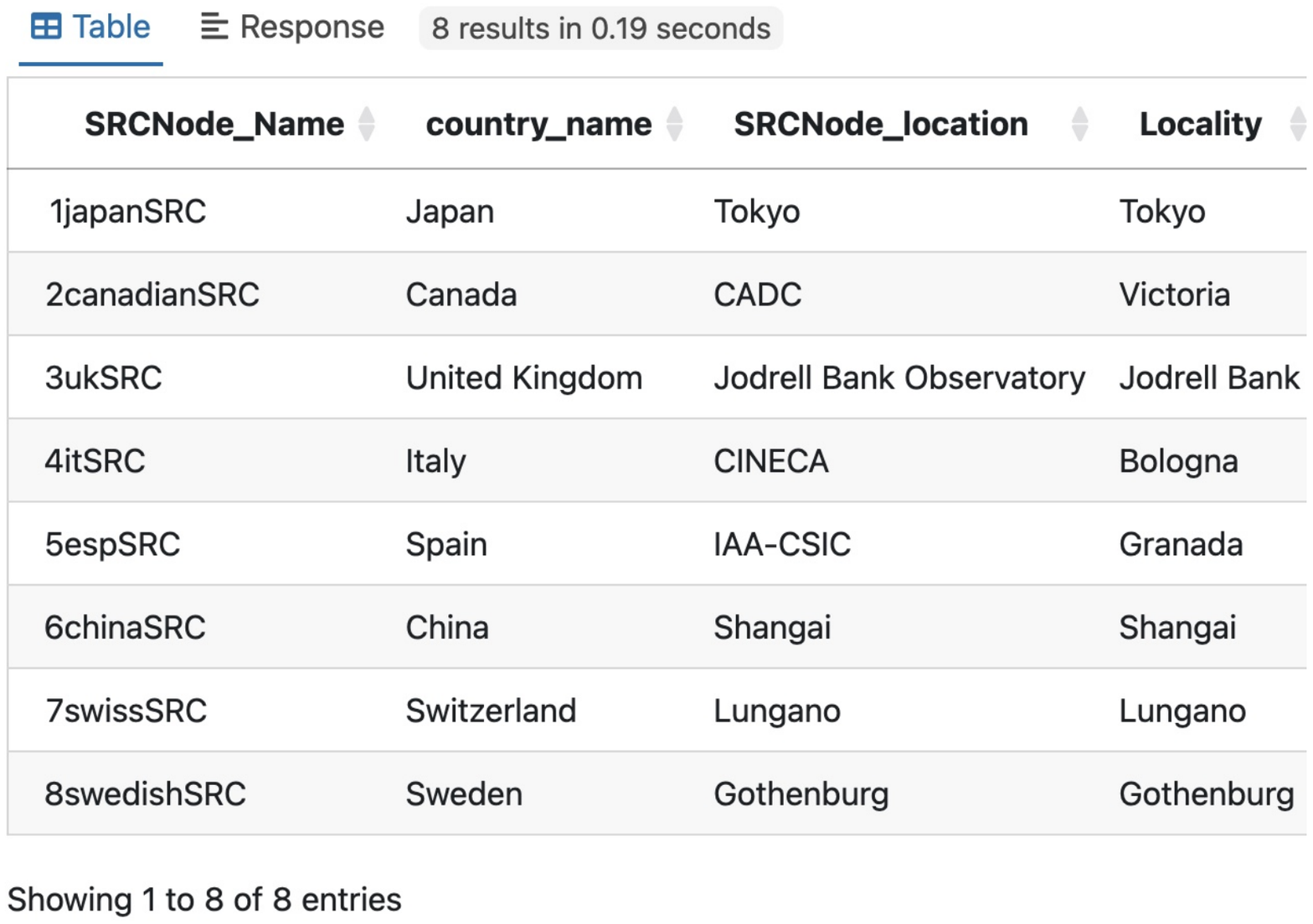}
    \caption{The list of the SRCs at the SRCNet 0.1} 
    \label{fig:answer1}
\end{figure}

As result of applying the previous instruction \ref{lst:sparQL1}, considering the TestData \cite{edgar202517608322}, we obtain as return, the following information \ref{fig:answer1}:

\captionsetup{justification=centering}

We can also request, through the following query \ref{lst:sparQL2}, for those nodes with the computing resources such as memory and number of cores in a certain domain.

\begin{lstlisting}[style=jsonstyle, caption={SPARQL query to obtain computing resources},label={lst:sparQL2}]

PREFIX aschema: <https://schema.ld.admin.ch/>
PREFIX ns0:     <http://example.org/>
PREFIX rdf:     <http://www.w3.org/1999/02/22-rdf-syntax-ns#>
PREFIX schema:  <http://schema.org/>
PREFIX schemas: <https://schema.org/>
PREFIX xsd:     <http://www.w3.org/2001/XMLSchema#>
PREFIX qudt: <http://qudt.org/vocab/unit/>

SELECT ?SRCnode ?numCores?memSize?Unit?numOfpflops
WHERE {
  ?SRCnode a ns0:Computing .
  ?SRCnode schema:hasCPU ?cores .
  ?cores  schema:cores?numCores.
  
  ?SRCnode schema:hasMemory?Memory.
  ?Memory a schema:QuantityValue;
        schema:value ?memSize ;
        schema:unit ?Unit .
   
  FILTER(?numCores > 44 && ?memSize < 20480)
}

\end{lstlisting}

By introducing the previous query \ref{lst:sparQL1}, we get as return, the following information \ref{lst:sparQL2}:

\captionsetup{justification=centering}
\begin{figure}[htbp]
    \centering
    \includegraphics[width=0.5\textwidth]{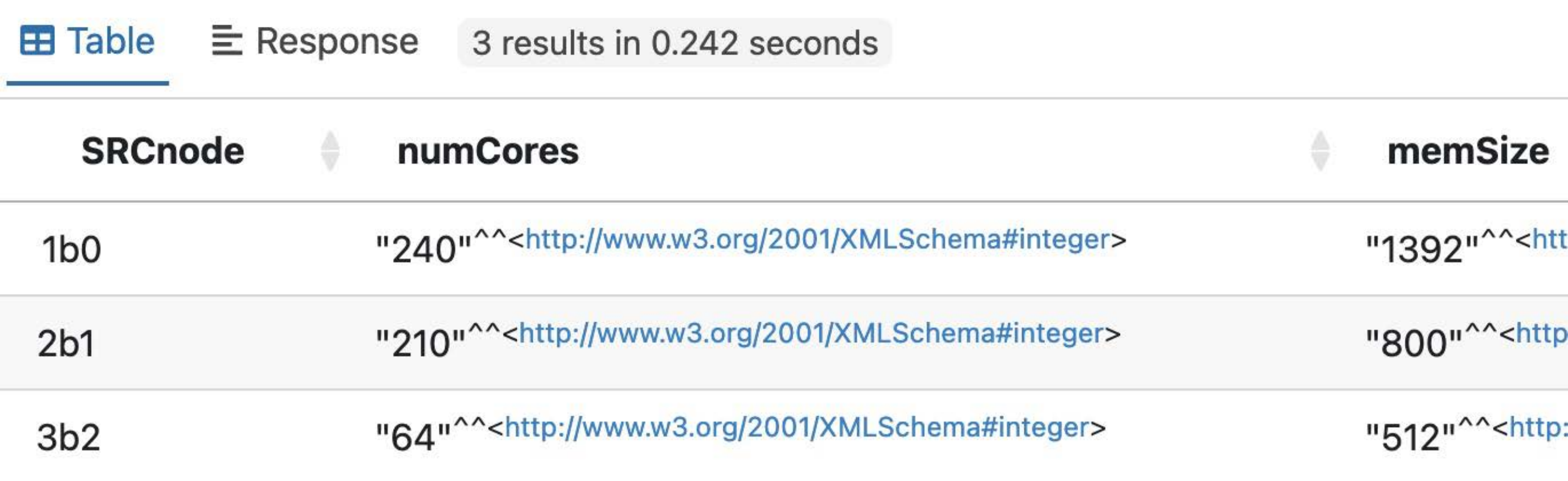}
    \caption{The list of the SRCs required}
    \label{fig:answer2}
\end{figure}

Similarly to the previous query,  we can access to the system and ask for those nodes with determined disk features. Therefore, it is important to remark that the requests will be done according to the required resources necessary to execute an specific operation, analysis or service. Anyway, the examples executed above give us a general procedural to make SPARQL queries and obtain information from the graph in a very efficient way.

\subsection{Competency Questions (CQs)}
\label{subsec:competency questions (CQs)}

In the sequence of what we have previously done in the early beginning of this section \ref{sec:validation}, we want to emphasize that for this kind of research the literature \cite{uschold1996ontologies} strongly recommends the implementation of competency questions to specify the knowledge being entailed in the ontology. Considered as a robust procedure to validate a semantic schema, the \textit{competency questions} strategy consists in formulating a set of questions designed to assess if the model is aptly adjusted with the problem domain and retains the capability to address effectively the queries proposed for it. 

By populating the semantic model for the SKA Regional Centre Network (SRCNet) with test data—rather than real-world data—we conducted an evaluation of the global structure of the model as presented in this research. In the following we formulated ten competency questions aimed at assessing its functionality, and obtained the corresponding answers through SPARQL queries. The SPARQL queries related to these competency questions, along with their outputs, can be accessed in the repository associated with this paper \cite{edgar202517608322}.\\

    \textbf{Q1: Where is located the file wallaby-center.fits?}\\
    
    This query aims to obtain the precise location of wallaby-center.fits file in the SRCNet thereby identify the SRCs containing the mentioned file. The query implies to access the local services available in the node, particularly the Datalake and check if there is a file identified as \texttt{wallaby-center.fits}. The resulting output will consist of pairs of SRC names and their corresponding Wallaby file identifiers enabling users to understand which datasets are linked to each SRC. This relationship highlights the accessibility of the data for analysis and usage within the SRC framework. In general, this query emphasizes the connection between SRCs and their respective datasets.\\

    \textbf{Q2: Who is the designated contact point at the canSRC?}\\
   
    This query is, in general, designed to obtain the contact point names of each SRC belonging in the SRCNet. However, in this particular case the query is directed to the canadian SRC (caSRC) specifying the caSRC entity by applying a filter to target its contact point. Once it is implemented, the result shows the relationship between the SRC and its associated contact point, allowing retrieval of the contact point name. The telephone, email, and address information for the contact point may also be included in the query, if needed. Basically , the output will include the SRC name and the name of the associated contact point, facilitating communication with the specified SRC.\\

    \textbf{Q3: Are there any nodes within the SRCNet that possess a minimum of 200 cores and 40 TB of available storage?}\\

    The aim of this query is to capture information about the SRCs that meet specific hardware criteria, precisely those with at least 200 CPU cores and a disk size of 40 gigabytes. Constructing this query implies connecting the computing node and its associated entities, particularly the CPU and Disk components. A filter is applied to ensure that only SRCs with the specified features are included in the results, allowing a focused analysis of resource allocation.\\

    \textbf{Q4: Which nodes deliver visualisation services, such as Carta, that facilitate the loading of an existing file named "wallaby-center.fits"?}\\

    The aim of this query is to gather information regarding SRCs and their associated datasets along with visualisation tools. The query identifies SRC nodes, extracting their names, Wallaby file identifiers, and the names of visualisation tools linked to the local service catalogue. It is structured to delineate relationships between SRCs, their local service catalogues, and the datasets associated with them, as well as the visualisation services available within these catalogues. The expected output includes SRC names, corresponding Wallaby files, and visualization tool names, providing a comprehensive view of both data resources and visualisation capabilities associated with each SRC. By capturing this information, it enhances understanding of how SRCs support data visualisation efforts.\\

    \textbf{Q5: What services are available in the spanish SRC?}\\ 

    This query extracts information about the available services provided by the SRC, identified as espSRC. The query is structured to retrieve the SRC name, the names of local services offered, service IDs, and descriptions associated with these services. It establishes relationships between the SRC node and its local service catalogue to access relevant details, applying a filter to restrict results to only those associated with espSRC. The expected output will include the SRC name, available service names, service IDs, and descriptions, providing insights into the specific offerings from espSRC and facilitating a better understanding of the services available.\\

    \textbf{Q6: Which SRCs contain multiple nodes within their infrastructure?}\\

    The aim of this query is to retrieve information about SRCs with multiple nodes along with their respective localities in a country (this query does not replace the scientific archive, which is the main mechanism for data discovery). It is structured using a subquery that identifies countries having more than one SRC node by grouping the results by country name and applying a count condition. The outer query then fetches SRC node names and localities for each country identified in the subquery. The relationships mapped include SRC nodes, their locations, and associated country names, ensuring that only relevant data concerning countries with multiple SRC nodes is displayed. The expected output consists of country names, SRC node names, and localities, ordered appropriately for clarity and insight into SRC distribution across identified countries.\\

    \textbf{Q7: Where can a jupyterNotebook be launched to execute a workflow on MeerKAT that requires more than 32 GB of RAM?}\\

    The query retrieves information about SRCs that provide jupyterNotebook services with specific memory requirements allowing to run a workflow on MeerKAT. The query identifies these SRC nodes and extracts their names alongside the names of Jupyter services offered, their corresponding memory size, and unit. It is structured to first establish the relationships between SRCs, local service catalogues, and Jupyter Notebooks, before accessing the resource requirements for those services. The filter condition ensures that only those Jupyter services with memory sizes exceeding 32 gigabytes are included in the results. The expected output will contain SRC names, associated Jupyter service names, memory sizes, and units, providing insights into the availability of computational resources linked to Jupyter services [REF to the output].\\

    \textbf{Q8: Which SRC possesses the largest number of CPUs?}\\
    
    This query aims to identify the SRC with the highest number of CPUs. The query is structured to extract SRC names and the corresponding number of CPUs by first establishing relationships between SRC nodes and their computing resources. It retrieves the SRC name and the number of cpus from the associated CPU properties. The results are ordered in descending order based on the number of CPUs, ensuring that the SRC with the greatest computational capacity is at the top. The query limits the output to a single result, effectively producing the SRC name that possesses the highest CPU number, which is crucial for understanding the computational resources available within the SRC framework.\\

    \textbf{Q9: What is the bandwidth available for the European nodes?}\\

     The query structures relationships between SRCs, their geographic locations, and their geopolitical classifications to extract relevant details. It gathers the SRC nodes belonging to Europe with their respective external bandwidth values, and its associated unit. By applying a filter, the query ensures that only SRCs from Europe and European Union are included in the results. The expected output is a list with SRC names along with their country, geopolitical classification, bandwidth values, and corresponding units, providing insights into the network capabilities and classifications of SRCs within relevant European contexts.\\

    \textbf{Q10: What type of disk is the spanish SRC using in its computing nodes?}\\ 

    This query retrieves the type of disk associated with the spanish SRC named espSRC. It is structured to first identify SRC nodes and then access their computing resources, specifically focusing on the disk. The query extracts the SRC name and the corresponding disk type from the relevant properties. A filter condition is applied to ensure that only results for espSRC are included in the output. The expected result provides insight into the storage infrastructure of espSRC. This information can be crucial for understanding the resource allocation and capabilities of the SRC.

\section{Final considerations and future works}
\label{sec:final considerations and future works}

In this work we have designed a semantic model to represent the architecture, data distribution, and computing services of the SKA Regional Centre Network (SRCNet), enabling to describe explicitly the nodes, resources, relationships, and workflows, thereby establishing a foundation for interoperability, efficient resources management, and advanced services across the distributed network infrastructure. Competency questions made on the model through SPARQL queries have validated its construction and integrity showing capability of capturing a vast spectrum of queries which confirms that the model is now ready to be instantiated and implemented with real-world data. In this sense, the scope of this research has been comprehensively addressed since that the semantic model developed is accurately aligned with the specifications of the SRCNet, as an highly complex network system. 

The documentation associated with the semantic model—including the JSON-LD file, dataset, full model diagram, SPARQL queries, and their respective outputs—is hosted in a publicly accessible GitHub repository, adhering to the principles of Findability, Accessibility, Interoperability, and Reusability (FAIR).   

Therefore, the result of this work will serve as a foundational infrastructure for a future service broker, enabling the efficient planning of tasks, workflows, and resource allocation. Since that the semantic model  incorporates both static and dynamic components, allowing two scenarios: one where a central entity hosts the information, and another where data is stored at individual nodes. In the latter case, whenever the future broker is required to plan a workflow, it can query the nodes to assess current system availability and make informed decisions.

On the other hand, this research represents a contribution to the field of semantics, in particular within the Astronomy Community. To our knowledge, at the date of writing, despite the International Virtual Observatory Alliance (IVOA) being responsible for defining schemas, protocols and vocabularies, managed by the Data Models, Data Access Layers, and Semantics Working Groups (WGs), there has been no prior effort to address the semantic model specifically for the SRCNet.

Once established the semantic model with the dynamic components included and with the capability for efficient querying demonstrated, we are now well positioned to express mathematically the SRCNet. As future work, it will be crucial to obtain a precise mathematical model for dynamic simulations aiming to investigate several network states, including load on the nodes, computing and storage resources availability, internal and external bandwidth, and services availability during the time. Therefore, considering the nature of the SRCNet, its architecture and configuration suggest that it should be represented mathematically through graph structures as a fully connected graph to facilitate required analysis. 

For all reasons mentioned above, this research not only contributes to the advancement of semantic modeling in astronomy but also gives rise to future works into the operational dynamics of the SRCNet enhancing our understanding of data processing and management in the context of the astronomy.

\section*{\textbf{Acknowledgments}}

Edgar Ribeiro João acknowledges financial support from the grant CEX2021-001131-S funded by MICIU/AEI/ 10.13039/501100011033, from the grant PRE2022-104630. The authors acknowledge financial support from the grant TED2021-130231B-I00 funded by MICIU/AEI/ and by the European Union NextGenerationEU/PRTR, from the grants PID2021-123930OB-C21 and PID2024-155817OB-I00 funded by MICIU/AEI/ and by ERDF/EU, and the grant INFRA24023 (CSIC4SKA) funded by CSIC.

\bibliographystyle{elsarticle-num-names}
\bibliography{references}








\end{document}